\documentclass[12pt]{article}
\usepackage{epsfig}%
\newcommand{\be}{\begin{eqnarray}}
\newcommand{\ee}{\end{eqnarray}}
\newcommand{\ba}{\begin{array}}
\newcommand{\ea}{\end{array}}

\newcommand{\slya}[1]{#1\hspace{-.5em}/\hspace{.15em}}
\begin{document}

\begin{center}
{\Large \bf  A novel way to probe distribution amplitudes of neutral mesons in $e^+e^-$ annihilation.}\\
\vspace{0.35cm}
  N.A. Kivel$^{1,2}$, M. V. Polyakov$^{1,2}$\\
  \vspace{0.2cm}
  \footnotesize{
$^1$Institute for Theoretical Physics II, Ruhr-University, 44780 Bochum, Germany\\
 $^2$   Petersburg Nuclear Physics Institute, 188300 Gatchina, Russia
}

\vspace{0.35cm}

\end{center}

\begin{abstract}
We derive the amplitude for the process $e^+e^-\to \pi^0\pi^0$ at
large invariant energy.  The process goes through the two-photon
exchange and its amplitude is expressed in terms of the
convolution integral  which  depends on the  shape of the pion
distribution amplitude (DA) and the centre of mass scattering
angle. Remarkable feature of the integral is that it is very
sensitive to the end-point behaviour of the pion DA -- it starts
to diverge if pion DA nullifies at the end-point as $\sqrt x$ or
slower. That makes the $e^+e^-\to \pi^0\pi^0$ process unique probe
of the shape of the meson DAs. The estimated cross section is
rather small, for $\sqrt s = 3$~GeV it ranges from a fraction of
femtobarn (for the asymptotic DA) to couple of femtobarn (for the
Chernyak-Zhitnitsky DA). The observation of the process
$e^+e^-\to\pi^0\pi^0$ with the cross section higher as estimated
here would imply very unusual form of the pion DA, e.g. the flat
one. The derived amplitude can be easily generalized to other
processes like $e^+e^-\to \sigma\sigma, K_SK_S, \eta\eta,
\eta^\prime\eta, \pi^0 f_2$, etc.
\end{abstract}

\noindent
{\bf 1.}
Distribution amplitudes (DAs) of mesons are the basic
non-perturbative objects which describe longitudinal momentum
distributions in the lowest $q\bar q$ Fock component of the meson
light-cone wave function. The meson DA is defined as the matrix element of
the  QCD operator on the light-cone sandwiched between vacuum and the meson.
For instance, the pion\footnote{In this paper we shall discuss and present all calculations for the case of
the pions, the generalization for other mesons is trivial.} DA $\phi(z)$ is  defined as the
following matrix element:

\be
\label{pionDA}
\langle 0 | \bar d (n) \gamma_\mu n^\mu \gamma_5 [n, -n] u (-n)
| \pi^+ (P )\rangle
&=&  i \sqrt 2 f_\pi (n\cdot P)
\int_0^1 dx \; e^{i (2 x - 1) P\cdot n} \phi(x)
\label{phi_pion}
\ee
Here $f_\pi=92.4$~MeV is the pion decay constant, $n^\mu$ is a light--like 4--vector ($n^2 = 0$), and
\be
[n, -n] &\equiv& \mbox{P}\, \exp \left[
\int_{-1}^1 dt\; n^\mu A_\mu (t n)
\right]
\label{P_exp}
\ee
denotes the path--ordered exponential, required by
gauge invariance. The pion DA is normalized by $\int_0^1 dx\ \phi(x)=1$ and it possesses
the symmetry $\phi(1-x)=\phi(x)$.

The DA is the universal object which enters description of many hard {\it exclusive} processes, such as
$\gamma^*\gamma\to \pi^0$, $\gamma \gamma\to 2 \pi$, pion form-factor, etc \cite{pionery}. One of classical examples is
the pion form factor which can be measured in $e^+e^-\to \pi^+\pi^-$ processes.
In the leading order in $\alpha_{\rm em}$ the process
occurs through the one photon exchange.
The asymptotic of the amplitude at large invariant energy ($\sqrt s \gg \Lambda_{\rm QCD}$) of colliding leptons can be written, up to the corrections of the order $\alpha_s^2$, as:

\be
\label{ff}
A(e^+(k)e^-(k^\prime)\to \pi^+(p)\pi^-(p^\prime))=   \frac{\alpha_{\rm em}(e_u-e_{d})}{9}\
\frac{4\pi \alpha_s f_\pi^2}{s^2}\ \left|\int_0^1 dx \frac{\phi(x)}{x(1-x)} \right|^2\ \bar l (k^\prime)\slya{ P} l(k)\, .
\ee
Here $e_u, e_d$ are quark charges of the corresponding flavour in the
units of the positron charge and we introduced the relative pion momentum $P^\mu=(p-p^\prime)^\mu$.
The strong coupling constant  $\alpha_s$ appears
due to the hard gluon exchange needed to make quarks and antiquarks collinear to the final pions.

\noindent
{\bf 2.}
Now if we turn to the process $e^+e^-\to\pi^0\pi^0$
the one photon exchange does not
contribute to the amplitude as the final state has positive C-parity.
The leading contribution is due to the two-photon exchange, see Fig.~1.
At large collision energy the $\pi^0$s are produced favourably  by two
$q \bar q$ pairs flying along two different light-cone directions with
quark and antiquark having small relative transverse momentum,  $k_\perp^2/s \ll 1$.
The simplest diagram to create such pairs
is shown on the right side of Fig.~1
\footnote{Note that for production of neutral {\it vector} mesons there is another type of diagrams in which
two photons are transformed into the vector meson independently of each other.
Such case has been considered recently in Ref.~\cite{peskin} and we do not dwell on it here. }.

\begin{figure}[t]
\begin{centering}
\includegraphics[scale=0.7]{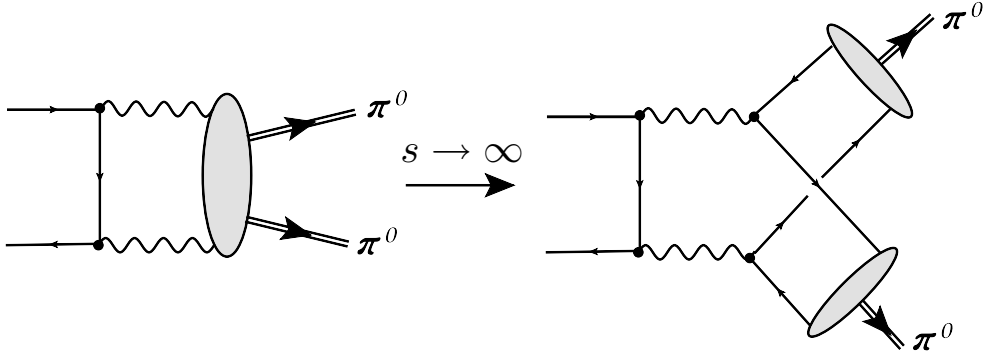}
\par\end{centering}
\caption{Two photon exchange diagram for the $e^+e^-\to\pi^0\pi^0$ process. On the right part of the figure the
leading QCD diagram is shown. There is another leading diagram with crossed photons which we do not show here.}
\label{fig1}
\end{figure}

An analysis of the diagram in Fig.~1 gives that the dominant virtuality of the intermediate photons is of
the order $\sim s$. That indicates
that the $q \bar q$ pairs are produced at a short distance and possible QCD corrections are suppressed
by small $\alpha_s$ at the scale of the order of invariant energy $\sim \sqrt s$.

Calculating the corresponding diagrams (see Fig.~1) we obtain the following result for the amplitude:

\be
\label{my}
\lim_{s\to\infty} A(e^+(k)e^-(k^\prime)\to \pi^0(p)\pi^0(p^\prime))=
  \frac{\alpha_{\rm em}^2(e_u^2+e_{d}^2)}{3}\
\frac{8 \pi^2  f_\pi^2}{s^2}\ G\left(\cos \theta_{\rm cm} \right) \bar l (k^\prime)\slya{ P} l(k) ,
\ee
where the dependence on the centre of mass scattering angle $\theta_{\rm cm}$ is
given by the function $G\left(\cos \theta_{\rm cm} \right)$.
This function is given by the following convolution integral of the pion DA:

\be
\label{fG}
G\left(z\right)=z \int_0^1 dx\int_0^1 dy \frac{\phi(x)}{x(1-x)}\ \frac{\phi(y)}{y(1-y)}
\frac{[x(1-x)+y(1-y)]}{[(x+y-2 xy)^2-z^2 (x-y)^2 ]}\, .
\ee
This function is obviously odd in $\cos \theta_{\rm cm}$, as it should be on the basis of Bose symmetry.
Also we note that Eq.~(\ref{my}) is derived under the assumption
that $s\sin^2\theta_{\rm cm}\gg \Lambda_{\rm QCD}^2$, that corresponds to the kinematical range of the Mandelstam variables
$|t|\sim |u|\sim s\gg \Lambda_{\rm QCD}^2$. It is easy to see that
the convolution integral (\ref{fG}) is divergent
if $\sin\theta_{\rm cm}$ goes to zero.

Comparing the annihilation amplitude into $\pi^+\pi^-$ (\ref{ff}) with that for $\pi^0\pi^0$ (\ref{my}),
we see that the latter has suppression by $\alpha_{\rm em}\approx 1/137$,
however it has the same scaling behaviour in $1/s$.
The annihilation into neutral pions has no suppression by $\alpha_s$. The factor $\alpha_s$ in Eq.~(\ref{ff})
is due to the hard gluon exchange needed to make relative transverse momentum in each $q\bar q$ pairs producing
pions small, in the case of annihilation into $\pi^0\pi^0$ one does not need gluon exchange for that.
The expression (\ref{my}) with the replacement $(e^2_u+e_d^2)\to 8e_ue_d$ provides an $\alpha_{\rm em}$
correction for the process $e^+e^-\to \pi^+\pi^-$ and can be used for more precise extraction of the pion
e.m. form factor from the annihilation process. Similar to considered here, a mechanism for the $\alpha_{\rm em}$
corrections in the elastic lepton scattering on the nucleon has been considered recently in Refs.\cite{Kobush,KVdh}.

A remarkable property of the convolution integral (\ref{fG}) is that it is sensitive to the end point
behaviour of the pion DA. It is divergent if the pion DA $\phi(x)$ at small $x$ behaves as $\phi(x)\sim x^\alpha$
with $\alpha\leq \frac 12$.
Similar feature is possessed by the convolution integrals which enter QCD description of the exclusive
charmonia decays \cite{CZ}.
All other known hard exclusive processes contain convolution integrals which diverge only if
the pion DA is not zero at the end points, i.e. if $\alpha\leq 0$. Recent measurements of the $\gamma^*\gamma\to \pi^0$
by the BaBar collaboration \cite{babar} show surprising rise of the scaled $\gamma\pi$ form factor at large $Q^2$.
In Refs.~\cite{tolya,ya} this phenomenon is explained by the flat pion DA. The measurement of the
$e^+e^-$ annihilation into two $\pi^0$s will allow to check this hypothesis due to the sensitivity
of the convolution integral (\ref{fG}) to the end-point behaviour of the pion DA. One would expect
the change of the scaling behaviour in $1/s$ for the cross-section -- slower than $1/s^3$.

\noindent {\bf 3.} The cross section of the process
$e^+e^-\to\pi^0\pi^0$ computed from the amplitude (\ref{my}) has
the form: \be \label{x-section} \frac{d\sigma}{d\cos\theta_{\rm
cm}}=\frac{25}{1458}\frac{\alpha_{\rm em}^4 \pi^3 f_\pi^4}{s^3}
\sin^2\theta_{\rm cm}|G(\cos\theta_{\rm cm})|^2. \ee Here the
function $G(\cos\theta_{\rm cm})$ is given by the convolution
integral (\ref{fG}) and depends on the form of the pion DA. In
order to show the sensitivity of the cross section
(\ref{x-section}) to the pion DA,  we plotted  in Fig.~2 the cross
section as the function of the scattering angle at the invariant
energy $\sqrt s=3$~GeV for three choices of the pion DA: the
non-relativistic one $\phi(x)=\delta(x-\frac 12)$, the asymptotic
one $\phi(x)=6x(1-x)$ and for the pion DA a'la Chernyak-Zhitnitsky
(CZ) $\phi(x)=30 x(1-x)(2 x-1)^2$ \cite{CZ}. As it is expected the
cross-section depends very strongly on the form of the pion DA,
however the cross section is very small-- it ranges from a
fraction of femtobarn for the asymptotic DA to a couple of
femtobarns for the CZ DA. It is challenge to measure such small
cross-section, also an observation of the $e^+e^-\to \pi^0\pi^0$
process with considerably higher cross section (say, at the level
of 10th of femtobarn at $\sqrt s=3$~GeV) would signal about very
unusual pion DA, e.g. the flat one. In Fig~2 by the vertical line
we show the range of angles at which the whole formalism can be
applied, it corresponds to the range of kinematical variable
$|t|,|u| \ge 1.5$~GeV$^2$ at $\sqrt s=3$~GeV.
\begin{figure}[t]
\begin{centering}
\includegraphics[scale=0.5]{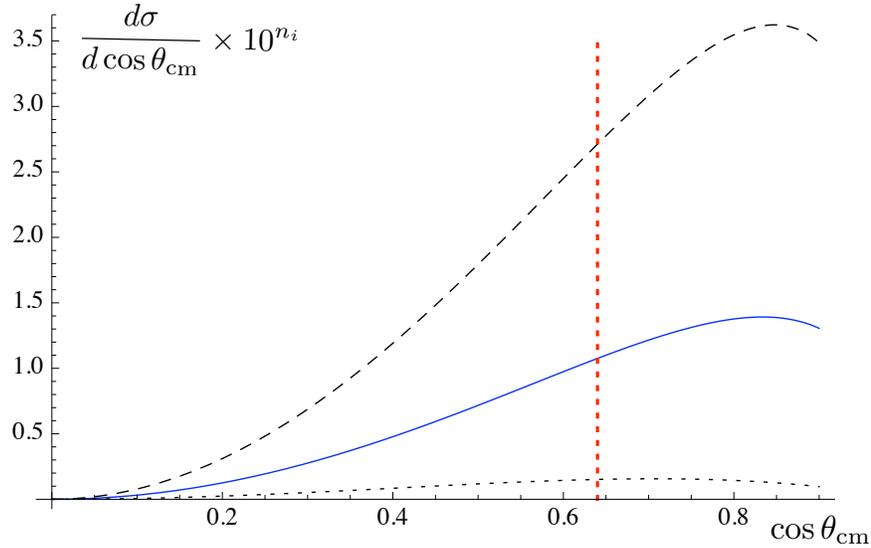}
\par\end{centering}
\caption{The angular dependence of the cross section in femtobarns
of the $e^+e^-\to \pi^0\pi^0$ process at $\sqrt s=3$~GeV. The
dashed line corresponds to the model for pion DA a`la
Chernyak-Zhitnitsky $\phi(x)=30 x(1-x)(2 x-1)^2$ (scale factor
exponent $n_{CZ}=0$), the solid line corresponds to the asymptotic
pion DA $\phi(x)=6x(1-x)$ (scale factor exponent $n_{asy}=1$), and
the dashed line with short dashes corresponds to the cross section
with non-relativistic pion DA $\phi(x)=\delta(x-\frac 12)$ (scale
factor exponent $n_{NR}=1$). By the vertical line we show the
range of angles at which the whole formalism can be applied, it
corresponds to the range of kinematical variable $|t|,|u| \ge
1.5$~GeV$^2$.} \label{fig2}
\end{figure}

Let us now to study the shape of the angular dependence of the cross section. For that we present
the cross section (\ref{x-section}) in the following form:

\be
\label{x-section-shape}
\frac{d\sigma}{d\cos\theta_{\rm cm}}=\frac{25}{1458}\frac{\alpha_{\rm em}^4 \pi^3 f_\pi^4}{s^3}
N\ S(\theta_{\rm cm}),
\ee
where the shape function $S(\theta_{\rm cm})$ and normalization constant $N$ are given
in terms of the convolution integral (\ref{fG}) as follows:

\be
\label{shape}
S(\theta_{\rm cm})=\frac{1}{N}\ \sin^2\theta_{\rm cm} |G(\cos\theta_{\rm cm})|^2,\ \ \
N=\int_{-1}^1d\left(\cos\theta_{\rm cm}\right)\sin^2\theta_{\rm cm} |G(\cos\theta_{\rm cm})|^2\, .
\ee
The total cross section is obviously expressed in terms of the normalization constant
$N$ as follows:
\be
\sigma_{\rm tot}=\frac{25}{1458}\frac{\alpha_{\rm em}^4 \pi^3 f_\pi^4}{s^3} N\, .
\ee

In order to study the dependence of the normalization and the shape function on the form
of the pion DA, let us choose simple one parametric Ansatz for the pion DA taking into account
only the first term in the Gegenbauer expansion:
\be
\label{param}
\phi(x)=6x(1-x)\left(1+a_2 C_2^{\frac 32}(2 x-1)\right),
\ee
where parameter $a_2$ we treat as a free. The result for the convolution integral (\ref{fG})
with the Ansatz (\ref{param}) is the following:
\be
\label{Ga2}
\frac{G(z)}{36}&=&\ln\left(\frac{1+z}{1-z}\right)+a_2 g_1(z)+a_2^2 g_2(z),\\
\nonumber
g_1(z)&=&\left(15 z^2-8\right) \ln \left(\frac{1+z}{1-z}\right)+\frac{15}{4} z
\left(\pi ^2 \left(1-z^2\right)+\left(1-z^2\right) \ln
   ^2\left(\frac{1+z}{1-z}\right)-4\right),\\
   \nonumber
   g_2(z)&=&
   -\frac{3}{4} \left(5 z \left(60 z^2+3 \pi ^2 \left(5 z^4-8 z^2+3\right)-56\right)+15 z \left(5 z^4-8 z^2+3\right) \ln
   ^2\left(\frac{1+z}{1-z}\right) \right.\\
   \nonumber
   &-&\left.4 \left(75 z^4-95 z^2+22\right) \ln \left(\frac{1+z}{1-z}\right)\right)
\ee
The result for the normalization constant $N$ in Eq.~(\ref{shape}) is the following:
\be
\frac{N}{(36)^2}&=&\frac{4}{9}(\pi^2-6)+\frac{8}{9}\left(\pi^2+3\right) a_2+
\frac{2}{63}\left(96\pi^4-800\pi^2-399\right) a_2^2 \\
\nonumber
&-&\frac{4}{77}\left(576\pi^4-6644\pi^2+8547\right) a_2^3+
\left(\frac{94464}{1001}\pi^4-\frac{7792}{7}\pi^2+1824\right) a_2^4\\
\nonumber
&\approx&1.72\ \left(1+6.65a_2+19.50a_2^2+27.76a_2^3+17.55 a_2^4 \right).
\ee
One sees that the normalization of the cross section has strong dependence on the parameter $a_2$.
This dependence is stronger than the corresponding dependence of the cross section for $e^+e^-\to\pi^+\pi^-$
reaction.
\begin{figure}[t]
\begin{centering}
\includegraphics[scale=0.5]{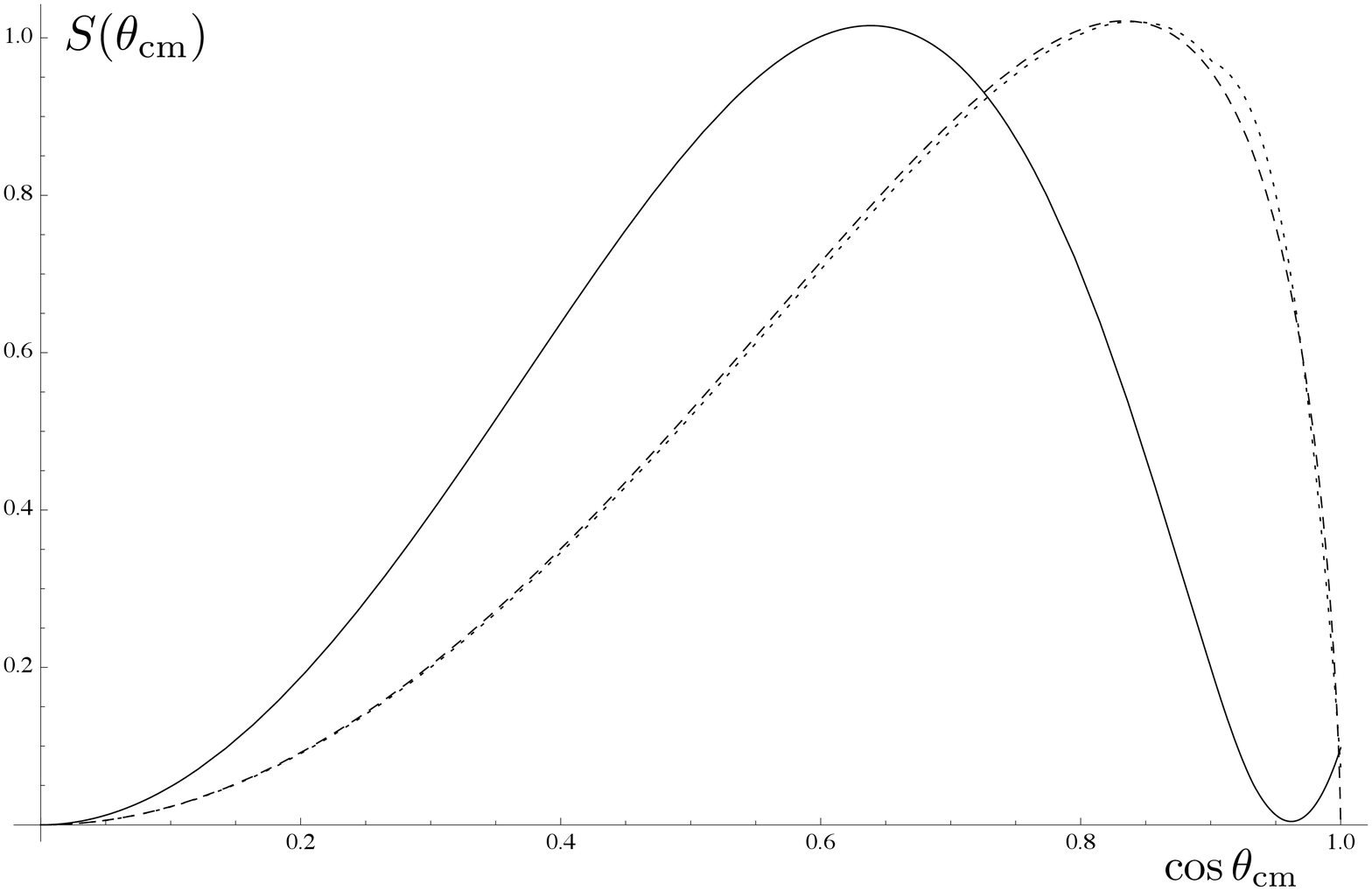}
\par\end{centering}
\caption{The shape function (\ref{shape}) as a function of the scattering angle. The dotted and dashed lines (almost indistinguishable )
correspond to the values of the parameter $a_2$ in Eq.~(\ref{param}) $a_2=0$ and $a_2=1$. The solid line corresponds to $a_2=-1/2$.}
\label{fig:multiplets}
\end{figure}

The shape function $S(\theta_{\rm cm})$ is shown in Fig.~3, where
we plot this function for several values of the parameter $a_2$. One sees that for positive
values of $a_2$ the shape function is almost indistinguishable from that for the
asymptotic pion DA. One can check that for the positive values of $a_2$ the corresponding shape functions can be hardly
distinguished from each other.
For the negative values of $a_2$ the shape function have different behaviour, see the solid line in Fig.~3.
That means that the measurements of the shape function would allows us to find out
if the pion DA is wider or narrower than the asymptotic one.
The knowledge of the shape function
does not allow to determine unambiguously pion DA. One can easily give examples of very different
pion DAs that leads to the same shape functions.

\noindent
{\bf 4.}
In summary, we derived the amplitude for the process $e^+e^-\to \pi^0\pi^0$ at large invariant
energy. The main result is presented by Eq.~(\ref{my}). The process goes through the two-photon
exchange and its amplitude is
expressed in terms of the convolution
integral (\ref{fG}) that depends on the  shape of the pion DA and cm scattering angle. Remarkable
feature of that integral is that it is very sensitive to the end-point behaviour of the pion DA -- it starts
to diverge if pion DA nullifies at the end-point as $\sqrt x$ or slower.
The estimated cross section is rather small, that is a challenge for the experiments. The observation of the process $e^+e^-\to\pi^0\pi^0$
with the cross section higher as estimated here would imply very unusual form of the pion DA, e.g. the flat one.
The derived amplitude
can be easily generalized to other processes like $e^+e^-\to \sigma\sigma, K_SK_S, \eta\eta, \eta^\prime\eta, \pi^0 f_2$, etc.

\section*{Acknowledgements}
Discussions with D.~Mueller, M.~Strikman, M. Vanderhaeghen and
A.~Vladimirov are greatly appreciated.
 This work was supported in parts by BMBF and
 by the Deutsche Forschungsgemeinschaft.


\end{document}